\documentclass[11pt]{article}
\usepackage[margin=1in]{geometry}
\usepackage{graphicx}
\usepackage{booktabs}
\usepackage{amsmath}
\usepackage{natbib}
\usepackage{hyperref}
\usepackage{xcolor}
\usepackage{caption}
\definecolor{linkblue}{RGB}{0,32,120}
\hypersetup{colorlinks=true, linkcolor=linkblue, citecolor=linkblue, urlcolor=linkblue}
\title{\textbf{Fine-Tuning an Order-Flow Foundation Model for Sub-Second Forecasting}}
\author{Fedor Shabashev}
\date{\today}

\begin{document}
\maketitle

\begin{abstract}
Generative models have become the standard approach in language and vision.
These models are typically pretrained first and then fine-tuned. Recent work
demonstrates how to pretrain an autoregressive generative model on Level-3
order-flow data. However, to our knowledge, no one has fine-tuned such a model yet for
downstream tasks.

Here we first pretrain a 300M-parameter generative model on Level-3 order-flow
data collected from two crypto exchanges, Bitstamp and Bitfinex. Then we
fine-tune the pretrained model to predict the microprice return and its
realized volatility. We show that our fine-tuned model outperforms classical
baselines, such as gradient boosting on handcrafted features, in predicting
microprice returns over sub-second horizons. Furthermore, the predictive
ability of the model transfers zero-shot to assets it was never fine-tuned on.
Due to the short-term nature of these forecasts, the predicted microprice moves
are smaller than the bid-ask spread plus fees, so the natural applications are
in execution and market making.

These results indicate that a pretrained generative model can be turned into a
competitive short-horizon price forecaster via fine-tuning.
\end{abstract}

\section{Introduction}
Level-3 market data is a sequence of orders that market participants send to
the exchange. Recent work \citep{tradefm2026arxiv,tradefm2026} shows that a generative model can
be trained to autoregressively predict the next event in the order sequence, one
event at a time. For most financial markets, including the crypto markets we
model here, the orders arrive at high frequency, often hundreds of orders per
second. At the same time submitting your own order to an exchange might take
more than 100\,ms and receiving data from the exchange takes tens of
milliseconds as well \citep{jha2020deep,albers2025latency,foley2023notsofast}. Hence a market participant will want to know the market prices hundreds of
milliseconds in advance, which is a much longer horizon than just one order
ahead. So a generative model predicting the
next order by itself cannot be used directly to answer practical questions such as ``is
it a good idea to submit a limit order at bid right now'', as by the time the
order arrives to the exchange it will be outside of the modeling horizon of
the generative model.

To convert the order flow into a form a neural network can consume, we
tokenize the orders \citep{tradefm2026arxiv,tradefm2026}. In our approach we normally convert
one order into one token, so ten buy limit orders will be converted into ten
tokens in the sequence. Crypto exchanges also allow their participants to
modify resting orders by changing the price or size. These update orders we
convert into two separate orders, a delete and an add, with the same
timestamp.

On the tokenized order flow, 2.8 billion training tokens, we train an
autoregressive decoder-only Transformer model with 300M parameters. The model
is trained to predict the next token with a cross-entropy loss.

We then take the generative model and fine-tune it via LoRA to predict the
first and second moments of the future microprice on the BTC/USD pair. The
microprice prediction is done for several horizons: 200\,ms, 500\,ms, 1, 2, 5,
10 and 30 seconds. For each of these horizons our model predicts the change in
microprice and its volatility. We then benchmark the fine-tuned model against
classical benchmarks such as gradient boosting and GARCH. We demonstrate that
our fine-tuned model outperforms all classical benchmarks in predicting the
microprice change direction at the 200\,ms and 500\,ms horizons, and
outperforms GARCH on volatility in four of five evaluation windows.
Furthermore, we demonstrate that the fine-tuned model can be
applied zero-shot to new assets it has not been fine-tuned on and is capable of
beating basic baselines there.

\paragraph{Contributions.}
\begin{enumerate}
  \item To our knowledge, the first fine-tuning of an autoregressive order-flow
  foundation model for predicting future price and volatility. We fine-tuned
  such a model and benchmarked it against classical baselines: order-book
  imbalance, order-flow imbalance, gradient boosting on those features, GARCH
  and persistence. We found that the fine-tuned model outperforms the
  baselines for predicting future price at sub-second horizons
  (Tables~\ref{tab:dir}--\ref{tab:gaps}).
  \item Zero-shot asset transfer. A generative model fine-tuned to predict the
  future microprice for BTC (Bitcoin) outperforms basic baselines when it
  comes to predicting the future microprice for SOL (Solana) and ETH
  (Ethereum).
  \item We found that the fine-tuned model performs well at sub-second
  horizons, but for horizons longer than a few seconds the classical methods
  match or beat it, which highlights the limitations of the current approach.
\end{enumerate}

\section{Related Work}\label{sec:related}
\paragraph{Supervised LOB forecasting.} A long line of work predicts
short-horizon price moves directly from the book with task-specific supervised
models. \citet{cont2014ofi} show that order-flow imbalance (OFI)---net signed
size at the best quotes---linearly predicts contemporaneous price change, and
\citet{kolm2023deepofi} extend multi-level OFI into a deep, multi-horizon alpha
signal; \citet{zhang2019deeplob} (DeepLOB) train convolutional-recurrent networks
on raw book snapshots for direction, and \citet{sirignano2019universal} show that
deep networks on LOB states learn price-formation features that are largely
universal across assets. Our directional baselines---multi-level
order-book imbalance and signed trade-flow, plus multi-level OFI at matched
lookbacks as the sequence-aware control---are the linear members of this
family. The distinction from our work is that these models are trained from
scratch on labeled targets, whereas we fine-tune a representation pretrained
generatively on unlabeled order flow.

\paragraph{Generative LOB models.} A separate thread models the \emph{distribution}
of order flow rather than a supervised target. \citet{nagy2023generative}
introduce a token-level autoregressive generative model of message flow;
\citet{tradefm2026arxiv,tradefm2026} (TradeFM) scale this into a foundation model with the
one-token-per-trade tokenization we adopt, and \citet{nagy2025lobbench}
(LOB-Bench) provide the K--S / $1$-Wasserstein marginal protocol we use to assess
generative fidelity. \citet{li2025mars} (MarS) build an order-level generative
foundation model of comparable ambition and do obtain forecasts from it---but
\emph{through simulation}: they roll the generator forward 128 times from each
starting point and aggregate the trajectories into a minute-scale up/down/flat
class, leaving the backbone's weights untouched and contrasting their approach
with ``task-specific forecasting'' precisely because no task-specific training
occurs. In our model that route would pay the price Section~\ref{sec:gap}
measures: an open-loop rollout accumulates the deep-sweep volatility error.
Reading the forecast off the representation instead, as we do, avoids it. These models
optimize distributional realism and, in the case of TradeFM, explicitly leave
task-specific fine-tuning as future work---the gap this paper addresses. \citet{linna2025lobert} (LOBERT) trained a 1.1M-parameter BERT-like encoder model
on four NASDAQ stocks and fine-tuned it for mid-price direction; the autoregressive
backbones above have not been fine-tuned. \citet{acero2026synthetic} pretrains
GPT-style models of up to 32M parameters on synthetic order flow from simulators
with known parameters and continues next-token training on one day of NASDAQ
messages, where the synthetic initialization beats from-scratch training on
held-out tickers; the objective remains next-token loss, so no forecasting task
is fine-tuned. The classical zero-intelligence generator
\citep{smith2003zi}, which we use as a baseline, Hawkes-process models of order
flow \citep{bacry2015hawkes}, and the canonical return stylized facts
\citep{cont2001} set what a faithful generator must reproduce.

\paragraph{Time-series foundation models.} Chronos \citep{ansari2024chronos},
TimesFM \citep{das2024timesfm}, and Moirai \citep{woo2024moirai} pretrain on broad
corpora of sampled time series and transfer zero-shot or with light fine-tuning.
They operate on regularly-sampled scalar or vector series, not event-level
Level-3 order flow, so adapting them to the LOB---where the signal lives in the
irregular \emph{sequence} of individual book events---is a different problem from
fine-tuning a native LOB generative model, which is what we study. This is the
sense in which the present fine-tuning is new: supervised LOB predictors are trained
from scratch, generative LOB models target distributional fidelity, and
time-series foundation models are not native to order flow.

\paragraph{Microstructure baselines and targets.} Our volatility baselines are
realized-volatility persistence and GARCH \citep{bollerslev1986garch} (HAR
\citep{corsi2009har} is the standard long-memory alternative, not run here); our forecast
target is the microprice, the high-frequency fair-value estimator of
\citet{stoikov2018microprice}. Together with OFI/imbalance for direction, these
are the baselines the fine-tuned model is measured against.

\section{Data and Tokenization}\label{sec:data}
\paragraph{Data.} We collect every Level-3 (L3) order-book event (adds, changes,
cancels, and trades) from two exchanges (Bitfinex, Bitstamp). The main model's
corpus spans roughly two weeks (20 September to 2 October 2026); the recency
control of Section~4 uses later, longer collections from the same feeds.

\begin{figure}[h]\centering
\includegraphics[width=0.62\textwidth]{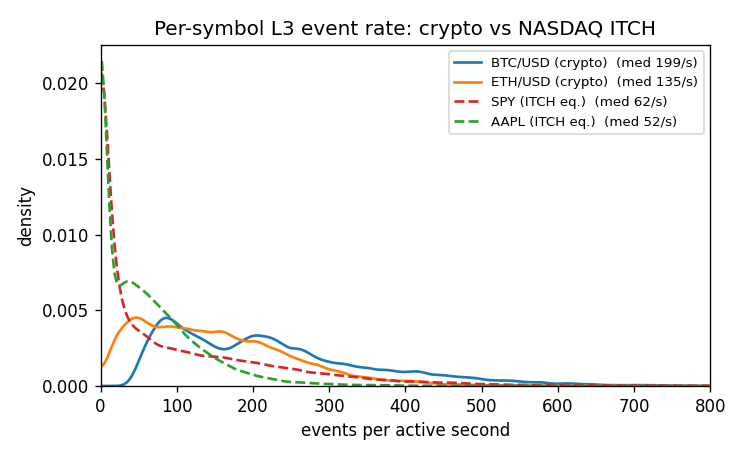}
\caption{Per-symbol event-rate density, crypto L3 vs NASDAQ ITCH equities (both
measured with our tokenizer's event definitions). Crypto (solid) sustains a
markedly higher per-second rate than even the busiest equities (dashed).}
\label{fig:rates}
\end{figure}

\paragraph{Event rates.} Crypto L3 is higher-frequency per symbol than the
equity ITCH feeds used by prior LOB models, and steadier (Fig.~\ref{fig:rates}).
On Bitstamp, BTC/USD runs a median of $199$ events/s; measured from NASDAQ
TotalView-ITCH 5.0 with the same event definitions, the busiest equities run
$3$--$4\times$ slower in the median (SPY $62$/s, AAPL $52$/s), though
spikier---single-second microbursts (AAPL $>$$28{,}000$/s) that crypto never shows (BTC max
$\sim$$1{,}000$/s). Two structural differences shape the tokenizer: crypto venues
stamp on a millisecond grid, collapsing $32\%$ of consecutive events onto an
identical timestamp (vs $5$--$8\%$ on ITCH's nanosecond grid), and trade
continuously rather than in a $6.5$\,h session.

\paragraph{Tokenizer.} Each add or cancel becomes one token with fields
(action, side, $\delta p$, $\log v$, $\Delta t$) at radices
$(2,2,16,16,16)$, a 16{,}384-element vocabulary, exactly as TradeFM. Trades do
not receive a token of their own: the stream is add/cancel only, the liquidity an
execution consumes appears as the removal of the resting order, and trades enter
the representation through the trade-based EW-VWAP mid against which $\delta p$
is measured. $\Delta t$ is the backward-looking inter-arrival time since
the preceding event. Crypto
required three departures from the equities recipe:
\begin{itemize}
  \item \textbf{Reserved $\Delta t{=}0$ bin.} Order activity is bursty at the
  millisecond: $32\%$ of consecutive events carry an identical
  timestamp---predominantly fleeting limit orders posted and canceled within one tick, plus
  orders batched in a single exchange message. We therefore reserve inter-arrival
  bin 0 for $\Delta t{=}0$ and quantile-bin only the strictly-positive tail into
  the remaining bins.
  \item \textbf{Reprice splitting.} A Bitfinex \texttt{change} (an order moving
  price) is emitted as a cancel of the old resting order followed by an add of
  the new one, so the token stream is pure add/cancel.
  \item \textbf{Pooled calibration, frozen edges.} Bin edges are fit once on a
  pooled sample of both venues and frozen, so a single token id denotes the
  same event across venues and across retrains.
\end{itemize}
Continuous fields are scale-invariant ratios: $\delta p$ is signed basis points
from the current mid, $\log v$ is $\log(1+\text{size})$, and the mid is an
exponentially-weighted VWAP of executed trades: a running average of trade
prices, each weighted by its size, with a $30$\,s exponential half-life so
recent and larger trades dominate. (We use a trade-VWAP mid rather than the
book midpoint because it is robust to a one-sided or momentarily empty book.)
Table~\ref{tab:tokens} traces a run of consecutive BTC/USD events through the
tokenizer.

\begin{table}[h]\centering\small\setlength{\tabcolsep}{4pt}
\caption{Eight consecutive Bitstamp BTC/USD book events and the tokens they
produce (26~October 2026; EW-VWAP mid $79{,}932.45$). $\delta p$ is signed
basis points from the mid,
$\log v=\log(1+\text{size})$, and $\Delta t$ the gap since the preceding kept
event. The bin tuple is (action, side, $\delta p$, $\log v$, $\Delta t$) with
action $0$ = add, $1$ = cancel and side $0$ = bid, $1$ = ask, and the id is its
mixed-radix value $8192a+4096s+256p+16v+t$. The bid $10\%$ below the mid falls
in the lowest $\delta p$ bin; the sixth row lands in the reserved
$\Delta t{=}0$ bin. Bitstamp stamps on a millisecond grid, and the quantile
edges at $0.9999$ and $1.0002$\,ms bracket it, so a $1$\,ms gap lands in bin 4
or 5 depending on floating-point rounding of the timestamp difference; the
edges are frozen, so this redundancy is left in place.}
\label{tab:tokens}
\begin{tabular}{llrrrrrlr}
\toprule
event & side & price & size & $\Delta t$ (ms) & $\delta p$ (bp) & $\log v$ & $(a,s,p,v,t)$ & id \\
\midrule
add    & bid & 71,939.21 & 0.0014 & 3  & $-1000.0$ & 0.001 & (0,0,0,0,7)  & 7 \\
cancel & bid & 71,939.21 & 0.0014 & 13 & $-1000.0$ & 0.001 & (1,0,0,0,10) & 8202 \\
cancel & ask & 79,940.91 & 0.0625 & 1  & $+1.06$   & 0.061 & (1,1,8,2,4)  & 14372 \\
cancel & ask & 79,932.46 & 0.2085 & 4  & $+0.00$   & 0.189 & (1,1,7,2,7)  & 14119 \\
cancel & ask & 79,932.46 & 0.1000 & 1  & $+0.00$   & 0.095 & (1,1,7,2,5)  & 14117 \\
cancel & ask & 79,932.46 & 0.1000 & 0  & $+0.00$   & 0.095 & (1,1,7,2,0)  & 14112 \\
cancel & ask & 79,950.00 & 0.7149 & 1  & $+2.20$   & 0.539 & (1,1,8,3,4)  & 14388 \\
add    & ask & 79,958.00 & 0.1200 & 1  & $+3.20$   & 0.113 & (0,1,9,2,4)  & 6436 \\
\bottomrule
\end{tabular}
\end{table}

\section{Model and Training}
We train a stock \texttt{LlamaForCausalLM} \citep{touvron2023llama} (RoPE, GQA,
SwiGLU, RMSNorm), context 1024 tokens, our main model being 300M parameters
(hidden 1152, 18 layers), for 4 epochs over 2.76B training tokens. The held-out
split is a temporally-later chunk (412M tokens), scored under one fixed protocol
(500 batches, fixed seed) for all cross-model comparisons.

\begin{figure}[h]\centering
\includegraphics[width=0.9\textwidth]{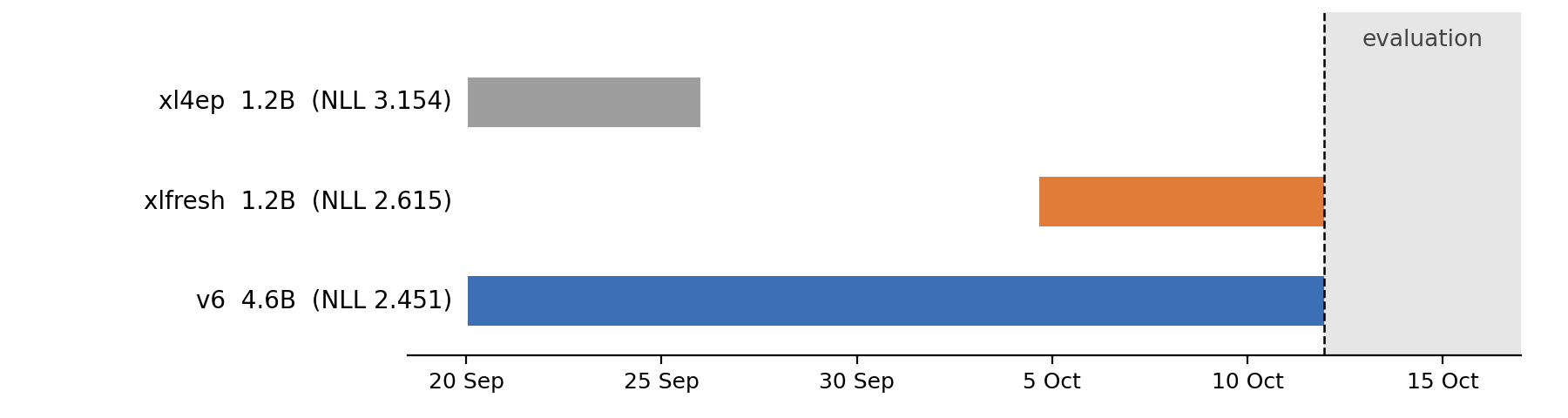}
\caption{Controlling for corpus recency. Three pretraining corpora on a date
axis. \emph{xl4ep} and \emph{xlfresh} have the same data budget and end 15 days
apart; \emph{xlfresh} and \emph{v6} end at the same instant and differ
$3.9\times$ in size. Each pair therefore varies one factor. All three are
scored in the same later period (shaded): next-token loss on a 12--13~October
chunk after the cutoff, and $0.2$\,s direction after LoRA fine-tuning on
13--15~October with the $20$k-step recipe of Section~\ref{sec:moments}, three
seeds per backbone, scored on the $33.5$-hour window of 15--16~October, which
is disjoint from the fine-tuning set and begins seven hours after it ends.
Recency moves the loss far more than size does ($-0.54$ vs $-0.16$ nats), but
size moves the downstream forecast more than recency does ($+0.012$ vs $+0.005$
correlation). One backbone pair per contrast. \emph{v6} serves only as the
control's large arm, not as a candidate main model: its corpus runs to
11~October and so overlaps the fine-tuning data and the 8 and 11~October
evaluation windows of Section~\ref{sec:moments} (the 5~October holdout is
excluded from it).}
\label{fig:recency}
\end{figure}

\paragraph{Scaling.} The regime is data-constrained (Table~\ref{tab:scaling}):
loss falls steadily with corpus size through 2.76B tokens, whereas enlarging the
model $1.53\times$ at fixed data (300M$\to$460M) buys only $-0.015$ nats. We do
not read a scaling law from the data column. The corpora grew by accretion, so
each larger corpus is also a more recent one and the holdout is later than all
of them, and a control that holds the corpus end date fixed while varying only
its size (Fig.~\ref{fig:recency}) finds that recency accounts for most of the
slope: on a later holdout, a corpus ending 15 days closer to it is worth
$\sim$0.5 nats at equal data budget, while $3.9\times$ more data at a fixed end
date is worth $\sim$0.16. Downstream the order reverses: the same two
contrasts move the fine-tuned model's $0.2$\,s directional correlation by $+0.005$
(recency) and $+0.012$ (size), one backbone pair per contrast and three
fine-tuning seeds per backbone (seed s.d.\ $0.003$--$0.007$; the size contrast
is seed-disjoint at all three sub-second horizons, the recency contrast only at
$0.2$\,s). Data still dominates model size in this range, by less than the raw
column suggests.

\begin{table}[h]\centering
\caption{Next-token loss (nats) on the held-out chunk, fixed protocol. Left: data
scaling at 300M. Right: the model-constrained test at fixed 2.76B.}
\label{tab:scaling}
\begin{tabular}{lcc}
\toprule
\multicolumn{2}{c}{\textbf{data scaling (300M)}} & \\
train tokens & NLL & \\
\midrule
1.23B & 2.742 & \\
1.63B & 2.647 & \\
2.26B & 2.452 & \\
2.76B & \textbf{2.309} & \\
\midrule
\multicolumn{2}{c}{\textbf{model scaling (2.76B)}} & \\
300M & 2.309 & \\
460M & 2.294 & ($-0.015$) \\
\bottomrule
\end{tabular}
\end{table}

\section{Generative fidelity}\label{sec:gap}
Before asking what the representation carries, we check that the pretrained
model is a faithful generator. We label the checks by market-data level: L2
denotes book-level quantities (spread, depth, imbalance) and L1 the price path
(returns), with L3 the order-flow events themselves. A rollout samples tokens autoregressively at
temperature 1; each token is de-quantized to an event and applied to a matching
engine, where a marketable add fills against the resting book and moves the
EW-VWAP mid, a non-marketable add rests, and a cancel removes the nearest
resting order (token ids carry no order identity, as in TradeFM). Feeding real
tokens through the same engine reconstructs a sensible two-sided book (spread
$\sim$1--2\,bps). Against the real continuation of the same seed windows, and
through the identical engine, we compare the model's rollouts and a
\emph{zero-intelligence} (ZI) stream that resamples each field independently
from its real marginal \citep{smith2003zi}, matching every marginal but
destroying joint and temporal structure. Following LOB-Bench
\citep{nagy2025lobbench} and TradeFM, distributions are compared by
Kolmogorov--Smirnov distance ($D_n$) and 1-Wasserstein ($W_1$), with no pairing
and no point forecast.

\paragraph{Book factors.} Table~\ref{tab:l2} shows the model matches the real
book's emergent factors as tightly as TradeFM's original equities model and
beats ZI on three of the four (spread, depth imbalance, resting-order
count---ZI's book collapses to 28 resting orders), ZI edging it only on touch
imbalance. On the joint structure of the token fields the model is
indistinguishable from a second real sample (Jensen--Shannon divergence at the
two-sample floor on three field joints, where ZI sits $5$--$9\times$ above it).

\begin{table}[h]\centering\small\setlength{\tabcolsep}{4pt}
\caption{L2 emergent factors, K-S distance to real (64 rollouts $\times$ 8192
events). TFM column is the original equities model for reference. ZI's thin
book is also lopsided (median depth imbalance $0.50$ against $0.00$ real),
which its K-S of $0.56$ reflects.}
\label{tab:l2}
\begin{tabular}{lcccccc}
\toprule
factor & median real & median model & median ZI & K-S model & K-S ZI & TFM (equities) \\
\midrule
spread (bps)            & 1.84  & 2.22   & 3.16  & \textbf{0.20} & 0.53 & 0.238 \\
touch imbalance         & 0.111 & 0.121  & 0.000 & 0.16 & 0.09 & 0.142 \\
depth imbalance (top-10)& 0.004 & $-0.008$ & 0.499 & \textbf{0.19} & 0.56 & 0.169 \\
resting order count     & 219   & 219    & 28    & \textbf{0.04} & 0.79 & --- \\
\bottomrule
\end{tabular}
\end{table}

\begin{figure}[h]\centering
\includegraphics[width=0.9\textwidth]{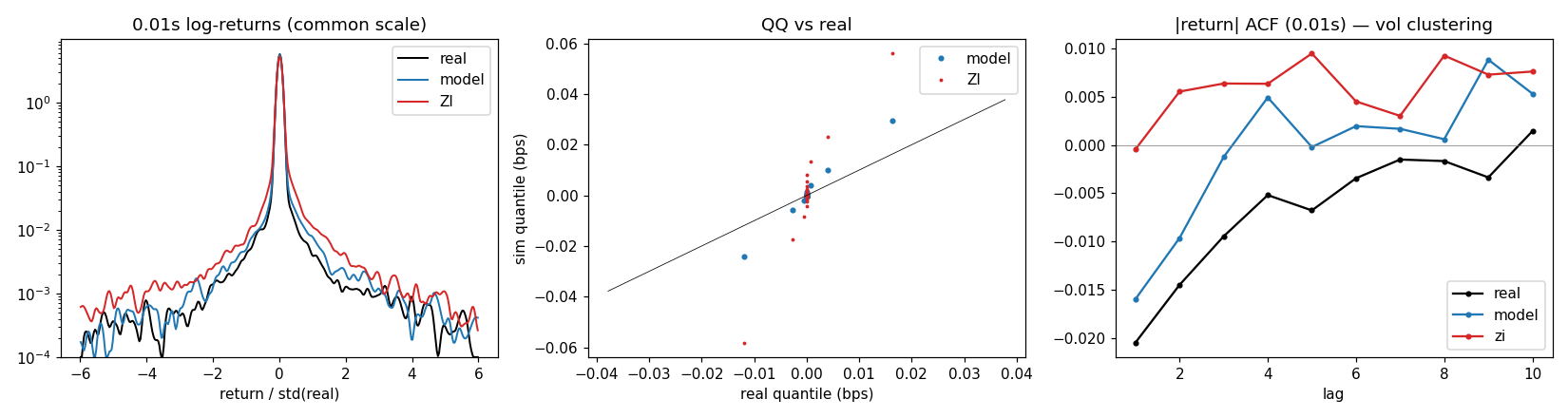}
\caption{Sub-second L1: return density on a common scale (all variants divided
by the real std), QQ vs real, and $|r|$-autocorrelation. Model (blue) tracks real
(black) in shape and reproduces volatility clustering; ZI (red) is visibly
fatter. The common scale prevents a single rollout's deep-sweep outlier from
rescaling an entire distribution.}
\label{fig:subsec}
\end{figure}

\paragraph{Returns and stylized facts.} The price process that emerges from the
model's flow reproduces the stylized facts---near-zero return autocorrelation
and slowly decaying $|r|$-autocorrelation---which ZI lacks entirely, and its
return distributions are closer to real than ZI's at every sub-second horizon
(K-S $0.02$--$0.06$ against $0.11$--$0.33$; Fig.~\ref{fig:subsec}). The typical
rollout is calibrated to within $\sim$1.2$\times$ of real volatility, but rare
rollouts in which persistent same-side flow sweeps $\sim$4$\times$ deeper into
the book than real flow does (median crossing depth $5.1$ vs $1.4$\,bps)
inflate the pooled std-based volatility ratio to $\sim$15$\times$, and the
inflation grows with rollout depth past about one context window
($\sim$5\,s). Two consequences matter for what follows. First, K-S misses this
tail entirely: most sub-second returns are an exact-zero atom, so the maximum
CDF gap sits at the mode, whereas $W_1$ registers it; we therefore report $W_1$
and realized volatility rather than K-S alone. Second, across five checkpoints
spanning $0.5$ nats of loss the deep-sweep frequency shows no monotone relation
to loss: next-token loss and open-loop price realism are decoupled, so reading
a forecast off the representation, as we do next, rather than off a rollout,
avoids a cost that scale does not remove.

\section{Fine-tuning for the first two moments}
\label{sec:moments}
We fine-tune the pretrained model to forecast, from the token context, the first
moment (signed mid return $r_h$; the mid is the microprice throughout, defined
below) and second moment (realized volatility $rv_h$)
of the near-future price across horizons from 200\,ms to 30\,s. The question is
whether the pretrained representation earns its keep against standard
microstructure baselines---so we benchmark, test for leakage, and measure
cross-asset transfer.

\paragraph{Supervised dataset.} We re-tokenize a slice of one asset
(bitstamp:btcusd) with the frozen edges---so tokens match the pretrained
corpus---and compute targets from the reconstructed \emph{microprice}
$(P_bQ_a{+}P_aQ_b)/(Q_a{+}Q_b)$ anchored at each token's \emph{true exchange
timestamp}. The book is rebuilt from L3 and \emph{re-seeded from a fresh snapshot
every $\sim$minute} so reconstruction error cannot accumulate; the reconstructed
microprice matches polled snapshots to $0.0$\,bps (median and p90 over sampled
instants). The microprice (not the sticky price-midpoint) is used because the BTC
touch price moves rarely, giving an otherwise-degenerate vol target.
$r_h=\log m(t{+}h)-\log m(t)$ and $rv_h=\sqrt{\sum(\Delta\log m)^2}$ over
$(t,t{+}h]$, with the base mid taken \emph{after} the current event (the model has
already seen it), so $r_h$ excludes the token's own price impact---conservative,
not leaky. An independent per-position recompute matches the vectorized targets to
$10^{-6}$\,bps and tokens re-tokenize deterministically.

\paragraph{Token-time alignment.} Each token's forward target must be anchored at
its true exchange time. Reconstructing token times by \emph{cumulative-summing}
the tokenizer's continuous $\Delta t$ is not equivalent: dropped warm-up and
invalid-price events advance the clock without emitting a token, so a cumsum clock
runs ahead with accumulating drift---here $\sim$2.25\,s, about $225\%$ of a 1\,s
horizon---which shifts every ``forward'' window into the token's own past and
overlaps its context, corrupting the target and inflating apparent skill. We
therefore emit the true per-token exchange timestamp and validate the resulting
targets by deterministic re-tokenization, an independent per-position recompute,
and snapshot ground truth. For an ultra-short-horizon forward target, how a token
is placed in time is load-bearing.

\paragraph{Head and transfer.} The last hidden state feeds a small head
$(\mu_h,\log\sigma_h)$ per horizon, trained with a robust decoupled loss
$\mathrm{Huber}(\mu,r)+\mathrm{Huber}(\log\sigma,\log rv)$. We use Huber rather
than a heteroscedastic Gaussian NLL because a naive NLL run collapsed the mean to
a near-constant, the documented failure of \citet{seitzer2022pitfalls} in which
the optimizer suppresses $\mu$ by inflating $\sigma$; the decoupled loss avoids
it by construction. We compare a
\emph{frozen probe} (head only) against \emph{LoRA} \citep{hu2022lora} (rank-16
attention adapters, 0.7\% of weights); the tables label the pretrained backbone
FM, for foundation model. LoRA beats the probe where there is signal
to extract, most at the shortest horizons ($+0.11$ direction correlation at
200\,ms versus $+0.01$ at 5\,s): adaptation surfaces signal the frozen features
do not, concentrated exactly where the model's edge lives.
Tables~\ref{tab:dir} and \ref{tab:vol} use $1{,}500$-step constant-LR fine-tunes;
the seeded analyses later in this section use the $10$k-step cosine recipe,
which reproduces the sub-second rows within the seed spread. LoRA: rank 16,
$\alpha{=}32$, dropout $0.05$ on q/k/v/o; batch 16, context 1024; AdamW with
learning rate $2\times10^{-4}$ (adapters) and $10^{-3}$ (head), weight decay
$10^{-4}$, gradient clipping at 1.0.

\paragraph{Protocol.} Holdouts are the last $10\%$ of each dataset by position
($\sim$3\,h on 5~October; 14:30--17:33 for the sub-second set), purged and
embargoed by the longest horizon the head forecasts. Confidence intervals use $15{,}000$ evenly spaced,
time-ordered holdout positions and a moving-block bootstrap
\citep{kunsch1989jackknife} with block length set to twice the longest horizon
divided by the median spacing of the sampled positions, floored at $25$ samples
(the floor binds for the sub-second and $1/2/5$\,s sets; the $5/10/30$\,s set
uses $108$). Model-minus-baseline
comparisons bootstrap the \emph{difference} of the two correlations on identical
resampled blocks: a paired bootstrap \citep{koehn2004statistical}, the analog
for a non-decomposable statistic of a Diebold--Mariano comparison
\citep{diebold1995comparing} in the spirit of White's reality check
\citep{white2000reality}.

The 200\,ms direction interval is the widest ($\pm0.06$) because the target is
extremely heavy-tailed: excess kurtosis is $\sim$180, and the top $5\%$ of $|r|$
windows carry $64\%$ of the $\mu$--$r$ covariance, so the Pearson correlation
inherits the sampling uncertainty of a few large-move blocks. The rank
correlation is three times tighter ($0.45\pm0.02$) and horizon-stable
($0.43$--$0.45$ at $0.2$--$1$\,s), and winsorizing the $1\%$ tails \emph{raises}
the point estimate to $0.51\pm0.02$: the rarest, largest moves are predicted
less well than the bulk. We keep Pearson as the headline because it is the
conventional, tail-inclusive statistic and the paired gaps are computed on it.

\begin{figure}[h]\centering
\includegraphics[width=0.95\textwidth]{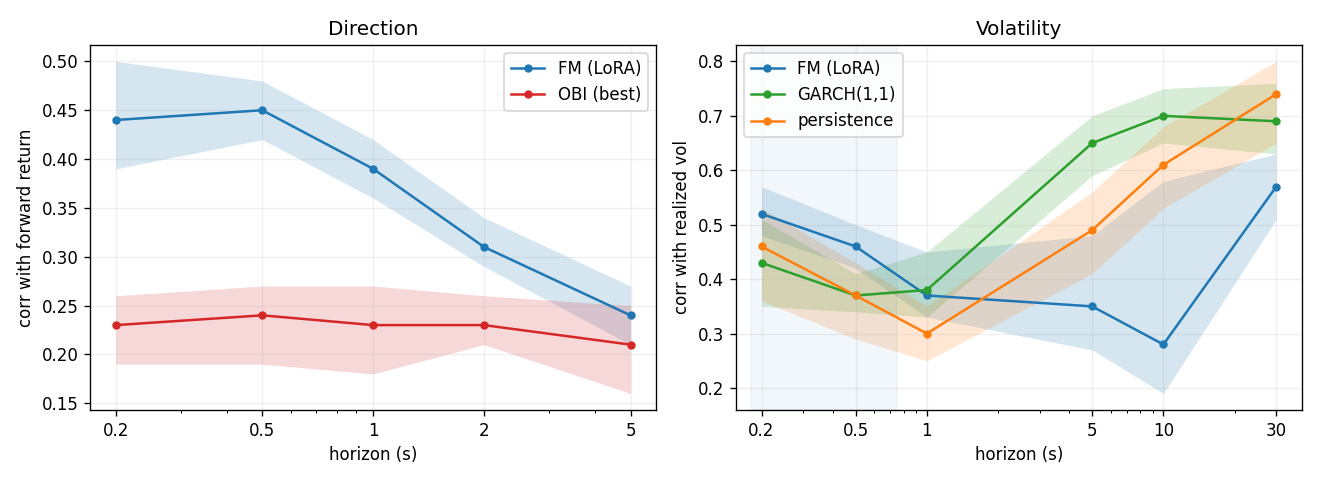}
\caption{Sub-second results. \emph{Direction} (left): the fine-tuned model
beats the strongest order-book-imbalance baseline at every horizon, and the gap
over this linear baseline widens toward the short end (over the gradient-boosted
baseline of Table~\ref{tab:dir} the gap is a steady $+0.05$--$0.08$; not
shown). \emph{Volatility} (right): the model beats GARCH and
persistence at 0.2--0.5\,s (shaded) but is overtaken by $\ge$5\,s, where the
sufficient statistic (recent realized vol) dominates. Classical signals decay
toward the ultra-short end, where the model's edge is largest. Direction points
at $0.2$--$1$\,s are head A and at $2$--$5$\,s head B; volatility points at
$0.2$--$1$\,s are head A and at $5$--$30$\,s head C, each head scored on its
own holdout window (Table~\ref{tab:heads}).}
\label{fig:subsecond}
\end{figure}

\paragraph{Sub-second results.} Tables~\ref{tab:dir} and \ref{tab:vol} and
Figure~\ref{fig:subsecond} benchmark on an embargoed holdout. On \emph{direction}
the model beats order-book imbalance (OBI), including a 10-level book and OBI
combined with recent-return momentum, at every sub-second horizon. The margin
grows toward the short end and plateaus near $+0.2$ correlation from $0.5$\,s
down to $200$\,ms. The $0.2$--$2$\,s gaps over OBI all exclude zero
(Table~\ref{tab:gaps} collects every paired gap); at 5\,s the model and OBI are
statistically tied ($+0.03$, CI $[-0.01,+0.08]$). A deeper book does \emph{not}
help: the 10-level imbalance is \emph{worse} than the touch at 200\,ms. Whatever the model reads is
not available in a richer snapshot of the book.

Beyond 5\,s the frozen representation ties OBI while the LoRA head degrades. At
$10$ and $30$\,s (Table~\ref{tab:dir}, head C) the paired LoRA-minus-OBI gaps are
$-0.07$ ($[-0.17,+0.04]$) and $-0.22$ ($[-0.44,+0.02]$), and the LoRA head's
directional skill is gone by $30$\,s ($-0.09\pm0.19$), whereas the frozen probe
($0.15/0.12$) still matches OBI ($0.17/0.13$). Beyond 5\,s, then, the adapter
stops adding to the frozen representation (on head C's window it is already
below the probe at 5\,s, $0.15$ vs $0.19$, Table~\ref{tab:heads}); these
long-horizon LoRA estimates are also the least precise in the table ($\pm0.10$,
$\pm0.19$), and we have not isolated the cause. It is not classical methods
reclaiming direction: the representation itself keeps pace with OBI.

On \emph{volatility} the same pattern holds but is thinner. The model beats
GARCH and trivial persistence at 0.2--0.5\,s by paired gaps of $+0.08/+0.09$
(GARCH) and $+0.06/+0.08$ (persistence), all CIs excluding zero, though the
0.2\,s persistence gap only marginally ($[0.00,0.16]$). At 1\,s the sub-second
head ties GARCH ($-0.00$, $[-0.04,+0.02]$) while still beating persistence
($+0.07$); the $1/2/5$\,s head already trails GARCH there ($-0.12$). From 2\,s
on GARCH wins significantly: $-0.17$ at 2\,s and $-0.09$ at 5\,s on head B's
window, and $-0.30/-0.42/-0.12$ at $5/10/30$\,s on head C's, all CIs excluding
zero. Persistence is the weaker opponent: on head B's window the model ties it
at 1 and 2\,s ($-0.01$, $+0.02$) and beats it at 5\,s ($+0.07$, $[0.02,0.12]$),
and only on head C's window does persistence win, by $-0.14/-0.33/-0.17$ at
$5/10/30$\,s, all CIs excluding zero. Beyond a second
or two, volatility clustering dominates: its sufficient statistic, recent
realized vol, is handed directly to GARCH but reaches the model only through
16-bin-quantized tokens.

These fine-tune correlations are not a seed artifact: re-run with a
cosine-decayed schedule over four random seeds, the sub-second LoRA reproduces
the values of Tables~\ref{tab:dir} and \ref{tab:vol} within a tight seed spread
(direction $\pm0.006$, volatility $\pm0.03$ at $0.2$--$1$\,s).

\paragraph{The target is the price, not the queue.} At 200\,ms the best bid or
ask moves in only $27\%$ of holdout windows ($35\%$ at 0.5\,s, $45\%$ at 1\,s), so
the plain mid $(P_b{+}P_a)/2$ is exactly unchanged $74\%/66\%/56\%$ of the time
and a microprice return in those windows is pure queue-size shifting. One might
therefore ask whether the sub-second correlation measures price prediction or
queue prediction. It is price prediction. Scoring the same $\mu_h$ against the
\emph{plain-mid} return gives correlation $0.52/0.50/0.42$ at
$0.2/0.5/1$\,s---\emph{higher} than against the microprice it was trained on
($0.44/0.45/0.39$)---and $82\%/83\%/75\%$ sign accuracy on the windows where the
price actually moves (correlation $0.58/0.55/0.46$ there). On the complementary
windows where the touch does not move, $\mu_h$ has essentially no skill against
the microprice change ($+0.07/-0.03/-0.09$): the model is not forecasting queue
shifts. Because the microprice is itself built from touch imbalance, OBI is
mechanically disadvantaged on that target; re-fitting OBI to the plain-mid return
raises it to $0.35/0.31/0.28$, yet the paired model-minus-OBI gap on the plain mid
is $+0.17/+0.19/+0.14$ (95\% CIs $[0.09,0.24]$, $[0.16,0.22]$, $[0.12,0.17]$),
indistinguishable from the paired microprice gap of Table~\ref{tab:gaps}
($+0.22/+0.21/+0.16$). The edge survives on the plain price.

\paragraph{The fine-tuned model beats gradient boosting on order-flow features.} OBI is a
snapshot; the natural classical \emph{flow} competitor is order-flow imbalance
\citep{cont2014ofi}---signed depth changes aggregated over a lookback---and its
multi-level extension \citep{kolm2023deepofi}. We reconstruct the book and fit
linear models on OFI at the top $K{=}10$ price levels over lookbacks matched to the
horizon, $(t-h,t]$ and $(t-5h,t]$, trained on the train split and scored on the same
embargoed holdout. The construction is sound---contemporaneous OFI--return
correlation is $+0.20$---but its \emph{predictive} power is weak: level-1 OFI alone
gives $0.07/0.04/0.00$ at $0.2/0.5/1$\,s, and ten-level OFI at the matched lookback
reaches $0.23/0.20/0.16$, the same level as the OBI snapshot ($0.23/0.24/0.23$);
longer lookbacks are worse ($0.14/0.11/0.09$). A linear flow aggregate therefore
captures what the snapshot captures and no more. A non-linear classical
baseline, a gradient-boosted model on the same features (ten-level OFI at both
lookbacks, OBI at one, five and ten levels, and the recent return), reaches
$0.39/0.37/0.31$, recovering most of the linear baselines' shortfall. The model's
paired gap over it is $+0.05/+0.07/+0.08$ (95\% CIs $[0.03,0.08]$, $[0.05,0.10]$,
$[0.05,0.10]$)---significant at every sub-second horizon, but roughly constant
rather than growing toward the short end. The large, horizon-growing margins over
OBI therefore partly measure how much a \emph{linear} snapshot leaves on the
table; against the best engineered-feature model, the fine-tuned model is
worth a steady $\sim$$+0.05$--$0.08$ in correlation.

\begin{table}[h]\centering\small\setlength{\tabcolsep}{3pt}
\caption{Direction: correlation of prediction with the forward mid return on the
embargoed holdout, as point $\pm$ half-width of a 95\% moving-block-bootstrap CI
(block bootstrap because overlapping forward windows autocorrelate the samples).
``Scratch (17M)'' is a same-tokens transformer trained from scratch, no
pretraining (mean of 3 seeds). OBI (best) is the strongest of touch / 5-level /
10-level imbalance and their combination with recent-return momentum, each fit
on train, the strongest chosen by holdout correlation (a choice that favors the
baseline); trade-flow is signed trade volume. GBM is a gradient-boosted model on 10-level OFI (lookbacks
$h$ and $5h$), OBI at 1/5/10 levels and the recent return (HistGradientBoosting,
300 iterations, learning rate 0.05, 31 leaves, min leaf 200, fit on 400k train
rows); computed for the sub-second set only. Bold marks the best point estimate
only; comparisons use the paired gaps of Table~\ref{tab:gaps}. The
\emph{head} column says which LoRA produced the row: A = co-trained on
$0.2/0.5/1$\,s, B = $1/2/5$\,s, C = $5/10/30$\,s, each on its own dataset and
embargoed holdout; Table~\ref{tab:heads} shows the sensitivity to this grouping.}
\label{tab:dir}
\begin{tabular}{llcccccc}
\toprule
corr$(\cdot,r)$ & head & OBI (best) & trade-flow & GBM & scratch (17M) & FM probe & FM LoRA \\
\midrule
0.2s & A & 0.23\,$\pm$\,0.04 & $-$0.12\,$\pm$\,0.13 & 0.39\,$\pm$\,0.04 & 0.19\,$\pm$\,0.03 & 0.33\,$\pm$\,0.04 & \textbf{0.44}\,$\pm$\,0.06 \\
0.5s & A & 0.24\,$\pm$\,0.04 & $-$0.06\,$\pm$\,0.07 & 0.37\,$\pm$\,0.03 & 0.20\,$\pm$\,0.03 & 0.38\,$\pm$\,0.03 & \textbf{0.45}\,$\pm$\,0.03 \\
1s   & A & 0.23\,$\pm$\,0.05 & $-$0.03\,$\pm$\,0.07 & 0.31\,$\pm$\,0.03 & 0.20\,$\pm$\,0.03 & 0.34\,$\pm$\,0.03 & \textbf{0.39}\,$\pm$\,0.03 \\
2s   & B & 0.23\,$\pm$\,0.03 & $-$0.03\,$\pm$\,0.04 & --- & 0.17\,$\pm$\,0.03 & 0.29\,$\pm$\,0.03 & \textbf{0.31}\,$\pm$\,0.03 \\
5s   & B & 0.21\,$\pm$\,0.05 & 0.04\,$\pm$\,0.08 & --- & 0.14\,$\pm$\,0.04 & 0.23\,$\pm$\,0.04 & 0.24\,$\pm$\,0.03 \\
\midrule
10s  & C & \textbf{0.17}\,$\pm$\,0.05 & 0.08\,$\pm$\,0.09 & --- & 0.09\,$\pm$\,0.05 & 0.15\,$\pm$\,0.05 & 0.10\,$\pm$\,0.10 \\
30s  & C & \textbf{0.13}\,$\pm$\,0.06 & 0.09\,$\pm$\,0.18 & --- & 0.07\,$\pm$\,0.07 & 0.12\,$\pm$\,0.05 & $-$0.09\,$\pm$\,0.19 \\
\bottomrule
\end{tabular}
\end{table}

\begin{table}[h]\centering\small\setlength{\tabcolsep}{4pt}
\caption{Volatility: correlation of prediction with realized vol on the embargoed
holdout (GARCH fit on train only, forecast out-of-sample), as point $\pm$
half-width of a 95\% moving-block-bootstrap CI. ``Scratch (17M)'' is the
from-scratch transformer (mean of 3 seeds; not scored for volatility on the
$1/2/5$\,s set). \emph{head}: A = LoRA co-trained on
$0.2/0.5/1$\,s, B = on $1/2/5$\,s, C = on $5/10/30$\,s, each on its own dataset and holdout
(Table~\ref{tab:heads}). The FM interval widens at long
horizons and is right-skewed at 10\,s, where few independent windows fit the
holdout ($^{\dagger}$: the 10\,s FM entry; full CI $[0.19,0.58]$); we do not
read the non-monotone $5/10/30$\,s FM values as a horizon effect. Paired
model-minus-baseline gaps (Table~\ref{tab:gaps}) are positive with CIs excluding zero at $0.2$--$0.5$\,s
(vs GARCH $+0.08/+0.09$; vs persistence $+0.06/+0.08$) and a tie vs GARCH at 1\,s on
head A; on head B GARCH wins at $1$--$5$\,s while persistence ties at $1$--$2$\,s
and loses at 5\,s; on head C both baselines win at $5/10/30$\,s.}
\label{tab:vol}
\begin{tabular}{llcccc}
\toprule
corr$(\cdot,rv)$ & head & persistence & GARCH(1,1) & scratch (17M) & FM LoRA \\
\midrule
0.2s & A & 0.46\,$\pm$\,0.09 & 0.43\,$\pm$\,0.08 & 0.31\,$\pm$\,0.06 & \textbf{0.52}\,$\pm$\,0.05 \\
0.5s & A & 0.37\,$\pm$\,0.07 & 0.37\,$\pm$\,0.04 & 0.30\,$\pm$\,0.06 & \textbf{0.46}\,$\pm$\,0.04 \\
1s   & A & 0.30\,$\pm$\,0.05 & \textbf{0.38}\,$\pm$\,0.06 & 0.29\,$\pm$\,0.05 & 0.37\,$\pm$\,0.06 \\
\midrule
1s   & B & 0.29\,$\pm$\,0.05 & \textbf{0.40}\,$\pm$\,0.05 & --- & 0.28\,$\pm$\,0.04 \\
2s   & B & 0.26\,$\pm$\,0.05 & \textbf{0.45}\,$\pm$\,0.05 & --- & 0.29\,$\pm$\,0.06 \\
5s   & B & 0.29\,$\pm$\,0.07 & \textbf{0.46}\,$\pm$\,0.06 & --- & 0.37\,$\pm$\,0.06 \\
\midrule
5s   & C & 0.49\,$\pm$\,0.07 & \textbf{0.65}\,$\pm$\,0.06 & 0.43\,$\pm$\,0.07 & 0.35\,$\pm$\,0.11 \\
10s  & C & 0.61\,$\pm$\,0.08 & \textbf{0.70}\,$\pm$\,0.05 & 0.51\,$\pm$\,0.07 & 0.28\,$\pm$\,0.20$^{\dagger}$ \\
30s  & C & \textbf{0.74}\,$\pm$\,0.08 & 0.69\,$\pm$\,0.07 & 0.54\,$\pm$\,0.07 & 0.57\,$\pm$\,0.06 \\
\bottomrule
\end{tabular}
\end{table}

\begin{table}[h]\centering\small\setlength{\tabcolsep}{4pt}
\caption{Sensitivity to the horizon grouping. The same horizon scored by two
different LoRA heads (each co-trained on its own horizon set, on its own dataset
and $\sim$3\,h embargoed holdout; the baselines move too because the holdout
window differs). Direction at 1\,s is stable; volatility at 1\,s and direction at
5\,s swing by $\sim$0.1---larger than the seed spread ($\pm0.006$ direction,
$\pm0.03$ volatility) and comparable to the claimed margins. Two confounds are
bundled here, the co-trained horizons and the scoring window; the paragraph
\emph{One adapter per horizon} below rules out the co-training, leaving the
window by elimination.}
\label{tab:heads}
\begin{tabular}{llcccccc}
\toprule
horizon & head & OBI & FM probe & FM LoRA (dir) & persistence & GARCH & FM LoRA (vol) \\
\midrule
1\,s & A ($0.2/0.5/1$) & 0.23 & 0.34 & 0.39 & 0.30 & 0.38 & 0.37 \\
1\,s & B ($1/2/5$)     & 0.24 & 0.35 & 0.38 & 0.29 & 0.40 & 0.28 \\
\midrule
5\,s & B ($1/2/5$)     & 0.21 & 0.23 & 0.24 & 0.29 & 0.46 & 0.37 \\
5\,s & C ($5/10/30$)   & 0.19 & 0.19 & 0.15 & 0.49 & 0.65 & 0.35 \\
\bottomrule
\end{tabular}
\end{table}

\begin{table}[h]\centering\small\setlength{\tabcolsep}{3pt}
\caption{Paired model-minus-baseline gaps in correlation (FM LoRA minus baseline)
with 95\% paired moving-block-bootstrap CIs on identical resampled blocks. The
marginal intervals of Tables~\ref{tab:dir}--\ref{tab:vol} overlap by
construction, since every predictor is scored on the same windows and shares
their block-to-block variation, so these paired gaps are the comparison
statistics: an interval excluding zero means the ranking is reliable on this
window, one spanning zero means a tie. $^{*}$: interval excludes zero ($^{\circ}$: marginal, lower
bound $0.000$). Head as in Table~\ref{tab:dir}; GBM was run for the sub-second set
only.}
\label{tab:gaps}
\begin{tabular}{llcccc}
\toprule
horizon & head & dir: FM$-$OBI & dir: FM$-$GBM & vol: FM$-$GARCH & vol: FM$-$persistence \\
\midrule
0.2s & A & $+0.22$ $[0.17,0.26]^{*}$ & $+0.05$ $[0.03,0.08]^{*}$ & $+0.08$ $[0.02,0.17]^{*}$ & $+0.06$ $[0.00,0.16]^{\circ}$ \\
0.5s & A & $+0.21$ $[0.18,0.24]^{*}$ & $+0.07$ $[0.05,0.10]^{*}$ & $+0.09$ $[0.05,0.12]^{*}$ & $+0.08$ $[0.02,0.18]^{*}$ \\
1s   & A & $+0.16$ $[0.14,0.20]^{*}$ & $+0.08$ $[0.05,0.10]^{*}$ & $-0.00$ $[-0.04,0.02]$ & $+0.07$ $[0.01,0.16]^{*}$ \\
\midrule
1s   & B & $+0.14$ $[0.12,0.17]^{*}$ & --- & $-0.12$ $[-0.16,-0.07]^{*}$ & $-0.01$ $[-0.06,0.06]$ \\
2s   & B & $+0.08$ $[0.05,0.10]^{*}$ & --- & $-0.17$ $[-0.23,-0.10]^{*}$ & $+0.02$ $[-0.02,0.09]$ \\
5s   & B & $+0.03$ $[-0.01,0.08]$ & --- & $-0.09$ $[-0.13,-0.06]^{*}$ & $+0.07$ $[0.02,0.12]^{*}$ \\
\midrule
5s   & C & $-0.03$ $[-0.12,0.06]$ & --- & $-0.30$ $[-0.37,-0.17]^{*}$ & $-0.14$ $[-0.22,-0.01]^{*}$ \\
10s  & C & $-0.07$ $[-0.17,0.04]$ & --- & $-0.42$ $[-0.52,-0.13]^{*}$ & $-0.33$ $[-0.43,-0.04]^{*}$ \\
30s  & C & $-0.22$ $[-0.44,0.02]$ & --- & $-0.12$ $[-0.20,-0.06]^{*}$ & $-0.17$ $[-0.22,-0.11]^{*}$ \\
\bottomrule
\end{tabular}
\end{table}

\begin{figure}[h]\centering
\includegraphics[width=0.6\textwidth]{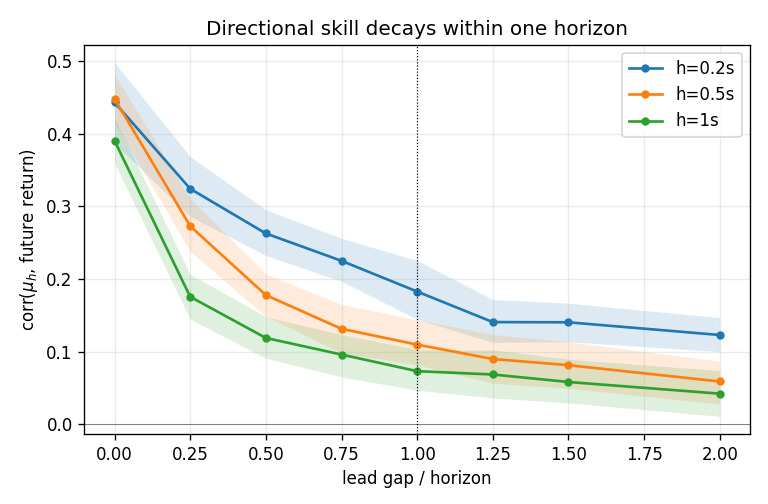}
\caption{Directional skill vs lead time: corr$(\mu_h,$ return over
$(t{+}\text{gap},t{+}\text{gap}{+}h])$ against lead gap (as a fraction of $h$),
with 95\% moving-block-bootstrap bands. gap$/h{=}0$ is the trained target; the
dotted line marks one full horizon of lead. The signal is a front-loaded,
short-lived continuation.}
\label{fig:decay}
\end{figure}

\paragraph{The edge is a short-horizon forecast.} We probe how far ahead the
signal reaches by sliding the realized window: we score $\mu_h$ against the return
over $(t{+}\text{gap},t{+}\text{gap}{+}h]$ as the lead gap grows
(Fig.~\ref{fig:decay}). The correlation is strongly front-loaded---it falls
$\sim$60--80\% within one horizon of lead and fades toward zero by two horizons,
leaving only a small residual (clearest at the shortest horizon). The edge
forecasts the \emph{imminent} order-flow continuation rather than a persistent
multi-horizon signal, which is why static snapshots and window-averaged scalar
features miss it.

\begin{figure}[h]\centering
\includegraphics[width=0.9\textwidth]{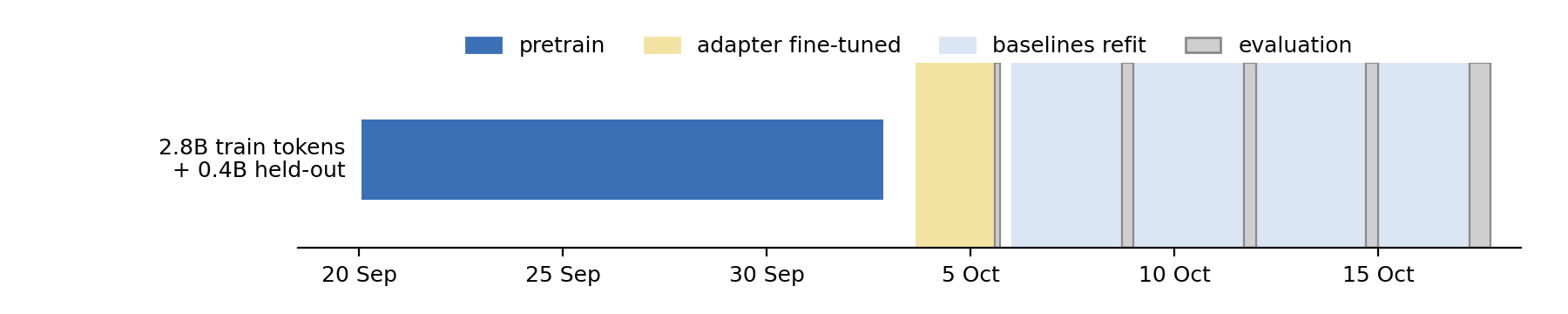}
\caption{Fine-tuning and evaluation on the time axis. The bar is the
backbone's pretraining corpus (training tokens plus the later held-out chunk),
which ends before the fine-tuning data begins.
The adapter is fine-tuned once, on 3--5~October (yellow), and the baselines are
fit on the same span; the 5~October holdout that follows is the first
evaluation window. Each later evaluation window of Table~\ref{tab:windows} is
the last $10\%$ of a three-day block by event count (so the quiet Saturday,
17~October, spans $12.5$\,h) whose first $90\%$ (blue) refits the
baselines alone; the adapter is carried forward unchanged. Model and baselines
are scored together on the gray windows.}
\label{fig:ftwindows}
\end{figure}

\paragraph{No leakage: fresh post-training data.} The strongest check against a
subtle leak is to score the trained model on data collected \emph{after} training
ends. On a dataset built from the
latest order flow---beginning 130\,min after the end of the fine-tuning dataset,
holdout included---the model gives direction
correlation $0.47/0.49/0.46$ and volatility correlation $0.52/0.48/0.43$ at
$0.2/0.5/1$\,s: \emph{at or above} the holdout. A leakage-inflated holdout would
collapse here; instead it is reproduced. (The fresh window is quieter, so
correlation rises while directional accuracy dips slightly---a regime effect, not
leakage.) A second, structural check: the pretraining corpus of the main model was
assembled on 2~October, before any of the fine-tuning data (3--5~October), so the
fine-tune holdout lies strictly outside the pretraining window and no holdout
sequence could have been memorized in pretraining (Fig.~\ref{fig:ftwindows}).

\paragraph{One adapter per horizon.} Table~\ref{tab:heads} leaves an ambiguity:
a head co-trained on $0.2/0.5/1$\,s and one co-trained on $1/2/5$\,s disagree
about $1$\,s by $\sim$0.1, but they also differ in the window they are scored on.
Either the co-training interferes, or the window does. We separate them by
training \emph{one adapter per target}---three LoRAs, each supervised on a single
horizon---on the same backbone, the same fine-tuning set and the same split (the
embargo is pinned to the joint runs' longest horizon, so the holdout is identical to
the token), with the $20$k-step cosine recipe and three seeds each, against a
joint adapter trained identically. All are scored on one further window of
$33.5$ hours (15--16~October), disjoint from every fine-tuning set and later than
every pretraining corpus used here.

Separating the horizons does not help; it costs a little. Direction falls from
$0.465$ to $0.453$ at $0.2$\,s (seed ranges disjoint), and from $0.487$ to
$0.483$ and $0.452$ to $0.449$ at $0.5$ and $1$\,s (overlapping); volatility is a
wash ($+0.006/-0.010/-0.006$, mixed signs). The loss is largest at the shortest
and noisiest horizon, which is where auxiliary supervision should help most: the
three targets are nested, so the longer ones act as a lower-variance teacher for
the structure the adapter must expose---the standard multi-task argument, in the
regime where it is expected to bite, since almost all of the adapter's capacity
is shared and only a two-output layer is task-specific
\citep{caruana1997multitask,baxter2000inductive}.

The useful consequence is negative: at most $0.012$, an order of magnitude below
the $\sim$0.1 swings of Table~\ref{tab:heads}. Per-horizon training therefore
\emph{cannot} account for those swings, which leaves the scoring window as the
explanation---consistent with the comparison across windows below, where absolute correlations
move by $0.03$--$0.05$ across windows while paired gaps stay put. We keep the
joint head.

\paragraph{Across evaluation windows.} A single $\sim$3-hour holdout cannot
distinguish a real edge from a favorable afternoon, so the comparison is run on
five holdouts: the 5~October window of Tables~\ref{tab:dir}--\ref{tab:gaps},
adjacent to the fine-tuning data, and four drawn from disjoint, later stretches
of the same instrument---8, 11, 14 and 17~October, of 7.1, 7.1, 7.5 and 12.5
hours, $34$ hours between them. Each window is built, split and embargoed
identically (last $10\%$ by position, purged by the longest horizon, $15{,}000$
evenly spaced positions, block bootstrap). Nothing about the model changes: the
\emph{same} adapter, fine-tuned once on 3--5~October, is carried forward to
windows up to twelve days later, while OBI, the gradient-boosted model, GARCH and
persistence are all \emph{refit} on each window's own training split. The
comparison is therefore stacked against the foundation model, which is stale by
construction while every baseline is local and current. We do not refit the
adapter per window: it costs a GPU fine-tune each time, and in preliminary
experiments refitting on fresher data brought no improvement. We leave a full
test to future work.

Table~\ref{tab:windows} reports the paired gaps. Direction holds without
exception: the model beats order-book imbalance in $15$ of $15$
window-horizon cells and the gradient-boosted baseline in $15$ of $15$, every
interval excluding zero. Against the non-linear baseline the margin on the four
later windows is in fact \emph{larger} than on 5~October---averaging
$+0.070/+0.113/+0.120$ against $+0.054/+0.074/+0.078$---so the figure
Tables~\ref{tab:dir}--\ref{tab:gaps} report is the conservative end of the
range, not its center. Signed trade-flow
remains uninformative everywhere ($15$ of $15$ intervals spanning zero).

The volatility picture is weaker and, more usefully, \emph{regime-conditional}:
the model beats GARCH at $0.2$ and $0.5$\,s on four of five windows but ties it
on 17~October ($+0.027$ and $+0.032$, both spanning zero), and the tie at
$1$\,s---reported as a null in Table~\ref{tab:gaps}---is itself the usual outcome
($4$ of $5$). We therefore weaken the second-moment claim: the model improves on
GARCH at the shortest horizons in most but not all market conditions.

\begin{table}[h]\centering\small\setlength{\tabcolsep}{4pt}
\caption{Paired gaps on five evaluation windows: the 5~October holdout of
Tables~\ref{tab:dir}--\ref{tab:gaps} and four later, disjoint holdouts
(7.1--12.5\,h each; 34\,h in total). The adapter is unchanged and fine-tuned
once on 3--5~October; every baseline is refit on each window's own training
split. Absolute correlations move across windows (the model's $0.2$\,s direction
$0.390$--$0.509$, OBI's $0.132$--$0.261$) while the paired gap stays within
$0.215$--$0.279$. $^{*}$: the 95\% paired moving-block-bootstrap interval excludes zero.}
\label{tab:windows}
\begin{tabular}{llccccc}
\toprule
gap & $h$ & 5~Oct & 8~Oct & 11~Oct & 14~Oct & 17~Oct \\
\midrule
FM $-$ OBI          & 0.2 & $+0.215^{*}$ & $+0.258^{*}$ & $+0.279^{*}$ & $+0.224^{*}$ & $+0.252^{*}$ \\
                    & 0.5 & $+0.213^{*}$ & $+0.240^{*}$ & $+0.250^{*}$ & $+0.202^{*}$ & $+0.231^{*}$ \\
                    & 1.0 & $+0.161^{*}$ & $+0.200^{*}$ & $+0.194^{*}$ & $+0.154^{*}$ & $+0.196^{*}$ \\
\midrule
FM $-$ GBM          & 0.2 & $+0.054^{*}$ & $+0.049^{*}$ & $+0.089^{*}$ & $+0.039^{*}$ & $+0.103^{*}$ \\
                    & 0.5 & $+0.074^{*}$ & $+0.103^{*}$ & $+0.143^{*}$ & $+0.078^{*}$ & $+0.128^{*}$ \\
                    & 1.0 & $+0.078^{*}$ & $+0.122^{*}$ & $+0.140^{*}$ & $+0.097^{*}$ & $+0.119^{*}$ \\
\midrule
FM $-$ GARCH        & 0.2 & $+0.084^{*}$ & $+0.119^{*}$ & $+0.115^{*}$ & $+0.075^{*}$ & $+0.027$ \\
                    & 0.5 & $+0.085^{*}$ & $+0.072^{*}$ & $+0.083^{*}$ & $+0.082^{*}$ & $+0.032$ \\
                    & 1.0 & $-0.004$ & $-0.004$ & $+0.044$ & $+0.025$ & $+0.059^{*}$ \\
\midrule
FM $-$ persistence  & 0.2 & $+0.061^{*}$ & $+0.103^{*}$ & $+0.175^{*}$ & $+0.041$ & $+0.183^{*}$ \\
                    & 0.5 & $+0.082^{*}$ & $+0.141^{*}$ & $+0.218^{*}$ & $+0.118^{*}$ & $+0.166^{*}$ \\
                    & 1.0 & $+0.074^{*}$ & $+0.102^{*}$ & $+0.167^{*}$ & $+0.089$ & $+0.166^{*}$ \\
\bottomrule
\end{tabular}
\end{table}

\paragraph{Zero-shot cross-asset transfer.} The backbone is pretrained on 73
asset streams across both venues (40 Bitstamp, 33 Bitfinex); ETH and SOL are among
them on both venues, so the \emph{representation} has seen their order flow, but
the LoRA and head are fine-tuned on BTC alone. Applied \emph{untouched} to Bitstamp
ETH/USD and SOL/USD (correlation is scale-invariant,
so the BTC-scale head is fair to compare), the BTC-trained model still beats each
asset's own baselines: direction correlation $0.26$--$0.33$ (ETH) and
$0.13$--$0.32$ (SOL) versus OBI $\le\!0.14$ and $\le\!0.05$ respectively, and
volatility correlation $\ge\!0.39$ on both. Direction is $\sim$40\% weaker than on
BTC, and the 200\,ms edge scales with liquidity (BTC $>$ ETH $>$ SOL), but at
0.5--1\,s the thin SOL book transfers as well as ETH. The fine-tuned mapping is
thus asset-general, with liquidity setting how short the horizon can be pushed---a
strong form of the foundation-model claim, since the model predicts assets it was
never fine-tuned for.

\paragraph{Does pretraining earn its keep?} We ablate pretraining directly by
training the \emph{same architecture family end-to-end from scratch}: 17M
parameters, random init, identical tokens and head, 3 seeds
(Tables~\ref{tab:dir},~\ref{tab:vol}). This gives a from-scratch model its best
shot, not just an adapter on frozen random weights. The control trains on the
\emph{identical} supervised split (same targets, same embargo, $\sim$24\,M
labeled tokens) with \emph{more} optimization than the LoRA fine-tune (up to
$8{,}000$ steps with early stopping vs.\ $1{,}500$), so it is not short-changed
on labels or compute. We size it at 17M because $\sim$24\,M labeled tokens is
already only $\sim$1.4 tokens per parameter; a 300M model trained end-to-end on
it would be far into memorization. A size sweep bears this out. From scratch,
direction correlation \emph{falls} with model size ($0.20/0.13/0.05$ at $1$\,s
for $17$/$75$/$300$M, all below OBI's $0.23$), and at the FM's own $300$M size it
collapses to chance while pretraining${+}$LoRA reaches $0.39$: identical
architecture and labels, pretraining the sole difference. What it lacks is the
$2.76$\,B \emph{unlabeled} pretraining tokens, the quantity under test.

On \emph{direction} the from-scratch model learns only about \emph{half} the
fine-tuned model's correlation at every horizon ($0.14$--$0.20$ vs
$0.24$--$0.45$): a real
signal, but most of it is a product of pretraining. It never beats even the simple
order-book-imbalance baseline (Table~\ref{tab:dir}: $0.14$--$0.20$ vs OBI's
$0.21$--$0.24$) and sits $\sim$0.2 below the gradient-boosted baseline
($0.39/0.37/0.31$), whereas the pretrained model clears that baseline at every
sub-second horizon. Pretraining is what moves the model from well below the
best engineered-feature baseline to slightly above it.

On \emph{volatility} the effect is horizon-dependent. Pretraining helps
sub-second ($0.31/0.30$ scratch vs $0.52/0.46$ at $0.2/0.5$\,s) but is
irrelevant beyond $5$\,s, where the from-scratch model \emph{matches or exceeds}
it ($0.43/0.51/0.54$ vs $0.35/0.28/0.57$), because long-horizon volatility is a
sufficient statistic learnable by direct supervision. Pretraining is thus
load-bearing for direction at every horizon and for volatility only below
$\sim$1\,s, concentrated exactly where the model wins. A cheaper control
(probe/LoRA on frozen random weights) agrees: random-init direction is
negligible (probe/LoRA corr $0.009$/$0.063$ at 1\,s on the $1/2/5$\,s set,
against $0.38$/$0.40$ for the pretrained probe/LoRA pair trained in the same
ablation run; the $1{,}500$-step runs of Table~\ref{tab:heads} give
$0.35$/$0.38$ for the same head and horizon).

\paragraph{Pretrain quality transfers to direction, not volatility.} We
fine-tune four backbones spanning a wide out-of-sample LM loss (NLL $2.94$--$3.43$
scored as a language model on the 5~October fine-tune holdout; this is a third
loss scale, since Table~\ref{tab:scaling} scores each model on the later
held-out chunk and Fig.~\ref{fig:recency} on a 12--13~October chunk, so only
orderings carry across them), all with pretraining corpora assembled before any
of the fine-tuning data, with an identical LoRA recipe (rank 16, cosine-decayed
schedule, $10$k steps). To separate
a genuine backbone effect from training-seed variance we run four seeds, with the
holdout held fixed, for the two endpoints: the weakest backbone and the
paper's main model.

Seed variance is sharply moment-dependent. Directional correlation is
reproducible to $\pm0.006$ across seeds, whereas volatility scatters by
$\pm0.03$, so single-run comparisons of pretrain quality sit below the noise.
With seeds, \emph{direction} shows a clean, statistically unambiguous transfer:
all four main-model seeds exceed all four weakest-backbone seeds at every
sub-second horizon ($+0.027/+0.019/+0.016$ correlation at $0.2/0.5/1$\,s, four
times the seed spread; one-sided permutation $p=1/70\approx0.014$, the smallest
value four seeds against four can give). Across the single-run curve
(0.2\,s) the gain is a step rather than a slope: the two weakest backbones sit
at $0.40$, the third at $0.42$, the main model at $0.43$. A fifth backbone (v5,
NLL $2.77$), excluded from the comparison because its corpus was assembled on
5~October and may contain the holdout, lands in the same place ($0.43$; seeded
$0.431$ vs $0.434$), so a backbone that may have seen the holdout gains nothing
from it: further evidence against leakage. \emph{Volatility} transfer is not
resolvable above the $\pm0.03$ seed scatter.

Better-pretrained backbones therefore give higher downstream directional
correlation. We do not read a scaling law from four mixed-corpus checkpoints; we
report only that \emph{better pretraining indicates better downstream direction}. Why the gain
is not larger has three candidate causes we cannot separate: a
near-efficient-market ceiling, limited pretraining scale, and a single-asset,
single-regime fine-tune set. Crypto order flow is not supply-limited the way
language corpora are, so two are directly testable (a fuller same-corpus
pretraining-scale study than the single control of Section~4, and a
multi-asset fine-tune), which we leave to future work. A \emph{frozen probe} does not capture even this: its downstream
correlation is flat in backbone quality, so a richer backbone is exploitable
through adapters but not through a linear read-out.

\paragraph{Predictable $\neq$ tradeable.} None of the sub-second forecasts clear a
$\sim$2\,bps one-way cost (half the $\sim$1--2\,bps spread plus a high-volume
taker fee of order 1\,bp; retail fee tiers are an order of magnitude higher, which
only strengthens the conclusion)---the moves are sub-basis-point. The signal is a real,
leakage-free, asset-general \emph{prediction}, but not directional alpha; its
plausible uses are in execution (anticipating imminent adverse selection) rather than taking.
This barrier is one of move \emph{magnitude}, not forecast accuracy: more or
more-diverse data could raise the correlation but not the size of a
sub-basis-point move, so the below-cost ceiling is unlikely to yield to scale even
as the prediction improves.

\paragraph{Inference latency.} An execution use is only meaningful if a forecast
arrives well inside the horizon it describes, so we measure the whole path from a
raw order-flow event to a prediction (Table~\ref{tab:latency}; batch 1, bf16,
context 1024, 200 timed repetitions after 30 warm-up, device-synchronized).
Tokenization---the C++ hot loop that turns one event into one token, excluding
file I/O, since a live feed delivers a parsed frame---costs $0.06$--$0.07\,\mu$s
per event, six orders of magnitude below the forward pass, and the moment head
costs $0.05$\,ms. The backbone is the whole budget. Two practical findings shape
it. First, the adapter is free once \emph{merged}: LoRA is an additive low-rank
update, so folding it into the base weights is exact, and leaving it unmerged
adds $\sim$144 small matrix products (four projections $\times$ 18 layers $\times$
two factors) worth $6$\,ms of pure kernel-launch overhead on the H100---$36\%$ of
total latency for $0.7\%$ of the weights. Second, batch-1 inference is
launch-bound rather than compute-bound: the 1024-token forward is $\sim$0.6\,TFLOP
($\sim$2\,ms of H100 compute) yet takes $10.4$\,ms eagerly, and on the H100 a full
re-encode and a single-token step against a KV cache cost the same ($10.4$ vs
$10.2$\,ms), which is only possible if neither is doing arithmetic. CUDA-graph
capture removes that overhead. We quote the \emph{full re-encode} under capture,
without a KV cache, because the two could not be combined in our implementation;
a streaming system does strictly less work per event, so the figure is an upper
bound on its cost. That figure---$4.0$\,ms, with the 99th
percentile $0.1$\,ms above the median---is $2\%$ of the shortest horizon we
forecast and sustains $247$ events/s against the $199$ events/s median rate of
this instrument, though not the $\sim$1{,}000 events/s bursts, where one would
forecast on a subsample. These numbers bound the model's own cost on one GPU;
they exclude a feed handler, an order gateway and the network, and so are not a
tick-to-trade figure.

\begin{table}[h]\centering\small
\caption{End-to-end inference latency, event to forecast. ``Merged'' folds the
LoRA into the base weights (exact); ``graphs'' is CUDA-graph capture via
\texttt{torch.compile(mode=\allowbreak"reduce-overhead")}. All rows re-encode the
full 1024-token window except the last, marked ``streaming'', which advances a KV
cache by one token; the two regimes cost the same on the H100 because both are
launch-bound there. Tokenization
($0.06$--$0.07\,\mu$s/event) and the head ($0.05$\,ms) are included in the
end-to-end column but are negligible against the backbone. Sustainable rate is
the reciprocal of the median.}
\label{tab:latency}
\begin{tabular}{llccc}
\toprule
configuration & device & p50 (ms) & p99 (ms) & events/s \\
\midrule
LoRA unmerged, eager        & H100 & 16.4 & 17.2 & 61 \\
merged, eager               & H100 & 10.4 & 11.8 & 96 \\
\textbf{merged + graphs}    & H100 & \textbf{4.0} & \textbf{4.1} & \textbf{247} \\
\midrule
LoRA unmerged, eager        & RTX 3090 & 25.5 & 30.3 & 39 \\
merged, eager (streaming)   & RTX 3090 & 12.4 & 18.2 & 81 \\
\bottomrule
\end{tabular}
\end{table}

\section{Discussion}
\paragraph{What the decoupling suggests.} Across our checkpoints, a lower
next-token loss buys a better event and book-factor generator but not a better
open-loop generator of price paths: the frequency of deep-sweep rollouts shows
no sign of falling as loss falls, while the same lower loss does transfer,
through adapters, to a better fine-tuned forecast (\S\ref{sec:moments}). The
representation improves with loss; the simulation does not. A plausible remedy
on the generative side is structural---feeding the emergent mid-price change
back into the context each step, a lever TradeFM uses and our tokens-only model
omits---and we leave it to future work.

\paragraph{Limitations.} Fine-tuning is on a single asset (bitstamp:btcusd) and
a single two-day window, though we test transfer zero-shot on two other assets;
and while the directional result rests on five disjoint holdouts totaling $37$
hours (Table~\ref{tab:windows}), all of it is one instrument on one venue
within a four-week stretch, so regime coverage is narrow. The corpus is $\sim$2 weeks (one market
regime), so the flat pretrain-to-probe transfer may be scale-limited rather than
intrinsic. Sub-second targets are discrete and the absolute margins modest
($\sim$+0.06 on vol); GARCH on 0.2\,s intervals is outside its native habitat,
and the comparison across windows shows the second-moment claim to be regime-conditional---the
model ties GARCH at $0.2$--$0.5$\,s on one of the five windows. At long horizons
the volatility comparison is regime-sensitive at the 3-hour-window scale: GARCH's
own 5\,s skill moves from $0.46$ to $0.65$ between two such windows
(Table~\ref{tab:heads}), so the size of its margin over the model should not be
read as a stable quantity. On the generative side, rollout volatility is tail-driven and run-to-run noisy, and we do not claim L1 parity with TradeFM. These remain exploratory
results, not a benchmark.

\paragraph{Future work.}
\begin{enumerate}
  \item Pretraining on more data. With a control for corpus recency, we
  established that more pretraining data leads to a more accurate fine-tuned
  model. At the moment we only pretrain
  on a few weeks' worth of data. Once we have collected more data, 100 days
  and more, we will retrain the models and test the resulting price and
  volatility prediction accuracy.
  \item Per-asset fine-tunes. Here we fine-tuned the pretrained model to
  predict the first two moments for the BTC/USD pair. We tested the resulting model on other assets
  and found that it beats basic baselines there. To understand how much
  accuracy we are leaving on the table by not fine-tuning per asset, we should
  test our zero-shot transfer accuracy against a per-asset fine-tune.
  \item Larger model. Our experiments indicate that a larger model has a
  modestly lower negative log-likelihood, hence scaling the model size might
  lead to a more accurate model and better downstream accuracy.
  \item State of the order book. The model we train here only looks at the
  order flow within the 1024-token context window. The model does not know
  the state of the order book at the beginning of the window. The state of
  the order book is potentially useful information, hence encoding it and
  feeding it to the model might improve the accuracy.
  \item A longer context window. For pairs with a lot of transactions, such as
  BTC/USD, a context window of 1024 tokens only covers a few seconds of the
  order flow. Since the model can only see 1024 tokens and nothing before
  that, we are truncating a lot of potentially useful information. Hence
  increasing the context window size might lead to higher accuracy.
  \item Modifications to the architecture. Linear attention or sliding-window
  attention might be helpful to save on compute once we have a longer context
  window.
\end{enumerate}

\section{Conclusion}
We have shown that a generative model pretrained to predict the next token in
Level-3 order flow can be fine-tuned to predict future microprice and
volatility. The fine-tuned model beats classical benchmarks in predicting the
microprice change direction at sub-second horizons and transfers zero-shot from
the asset it was fine-tuned on to others. Better pretraining transfers
downstream for direction: as the pretrained model gets better, the downstream
fine-tuned model gets more accurate as well, which suggests that a larger
pretraining dataset may improve the price prediction model further. All this
indicates that a next-token prediction model learns a representation that is
useful for predicting price at sub-second horizons.

\bibliographystyle{plainnat}
\bibliography{refs}
\end{document}